\documentclass[sigconf,nonacm,10pt]{acmart}
\usepackage{tabularx}
\usepackage{booktabs}
\usepackage{pifont}
\newcommand{\cmark}{\ding{51}} 
\newcommand{\xmark}{\ding{55}} 
\usepackage{array}
\usepackage{fancyhdr}
\usepackage{multirow} 

\usepackage{lipsum}
\usepackage[ruled,vlined]{algorithm2e} 
\usepackage{multirow}
\usepackage{url}
\usepackage[normalem]{ulem} 
\usepackage{epstopdf}
\usepackage{graphicx}

\usepackage{subfigure}
\AtBeginDocument{%
  }

\newcommand{\fref}[1]{\mbox{Figure~\ref{#1}}}

\newcommand{\myparab}[1]{\vspace{0.025in}\noindent\textbf{#1}}

\newcommand{\eg}{{\it e.g.}\xspace}
\newcommand{\ie}{{\it i.e.}}

\newcommand{\eat}[1]{}

\usepackage{color}

\usepackage{graphicx}
\usepackage{makecell}
\usepackage{caption} 
\usepackage{titlesec}
\titlespacing{\section}{0pt}{8pt plus 4pt minus 2pt}{2pt}
\titlespacing{\subsection}{0pt}{4pt plus 2pt minus 2pt}{2pt}
\titlespacing{\subsubsection}{0pt}{2pt plus 2pt}{2pt}

\begin{document}



\pagestyle{plain}     
\fancyhf{} 
\title{Simulating LLM training workloads for heterogeneous compute and network infrastructure}


\author{Sumit Kumar\textsuperscript{1},
Arjun Temura\textsuperscript{1},
Naman Sharma\textsuperscript{1},
Ramanjeet Singh\textsuperscript{1},
Meet Dadhania\textsuperscript{2},
Praveen Tammana\textsuperscript{2},
Satananda Burla\textsuperscript{3},
Abed Mohammad Kamaluddin\textsuperscript{4},
Rinku Shah\textsuperscript{1}}

\affiliation{%
  \textsuperscript{1}IIIT-Delhi \country{India}  \hspace{1em} 
  \textsuperscript{2}IIT Hyderabad \country{India} 
  \hspace{1em} %
  \textsuperscript{3}Marvell Technology Inc. \country{USA}
  \hspace{1em} %
  \textsuperscript{4}Marvell Technology Inc. \country{India} 
}

\renewcommand{\shortauthors}{}


\begin{abstract}
    The growing demand for large-scale GPU clusters in distributed model training presents a significant barrier to innovation, particularly in model optimization, performance tuning, and system-level enhancements. To address this challenge, LLM training simulators are employed to estimate training time and guide design decisions. However, the state-of-the-art LLM training simulators assume homogeneous compute and network infrastructure. In practice, device heterogeneity is inevitable due to resource sharing in cloud environments, frequent shifts in device generations, and inherent intra-chip interconnect heterogeneity. 
To address the gap between state-of-the-art and practical requirements, we propose the design of a heterogeneity-aware distributed LLM simulator capable of predicting training time while enabling abstractions to specify custom configurations for device groups \eat {parallelism groups,} and device-to-parallelism mapping. 
We present the design requirements and challenges in building a heterogeneity-aware distributed ML training simulator, and design components such as non-uniform workload partitioning. Our initial simulation results demonstrate the impact of heterogeneity on the model computation and communication time.
\end{abstract}

\maketitle


\section{Introduction} \label{sec:intro}

In the past decade, the emergence of transformer models, commonly referred to as large language models (LLMs), such as GPT~\cite{radford2019language}, Llama~\cite{touvron2023llama}, and Mixtral~\cite{jiang2024mixtral}, has revolutionized multi-task learning across various applications such as translation, summarization, and question answering. 
Despite this remarkable achievement, training the LLM comes with considerable costs of tens of thousands of GPUs for sustained operations over days and months. For example, Meta's Llama4 with 2 trillion parameters and 288 billion active parameters was trained on 32K GPUs~\cite{llama4}.
Given the high frequency of new model releases ({\em fifteen} per month on average~\cite{Integrated-AI}) and the massive resource demand to train these models, distributed ML training simulators are crucial for effective cost and capacity planning. The state-of-the-art simulators, ASTRA-sim~\cite{won2023astra} and SimAI~\cite{wang2025simai}, provide a full-system simulation framework for training clusters with homogeneous compute and networking infrastructure. 

\begin{figure}[t]
    \centering
    \includegraphics[width=0.4\textwidth]{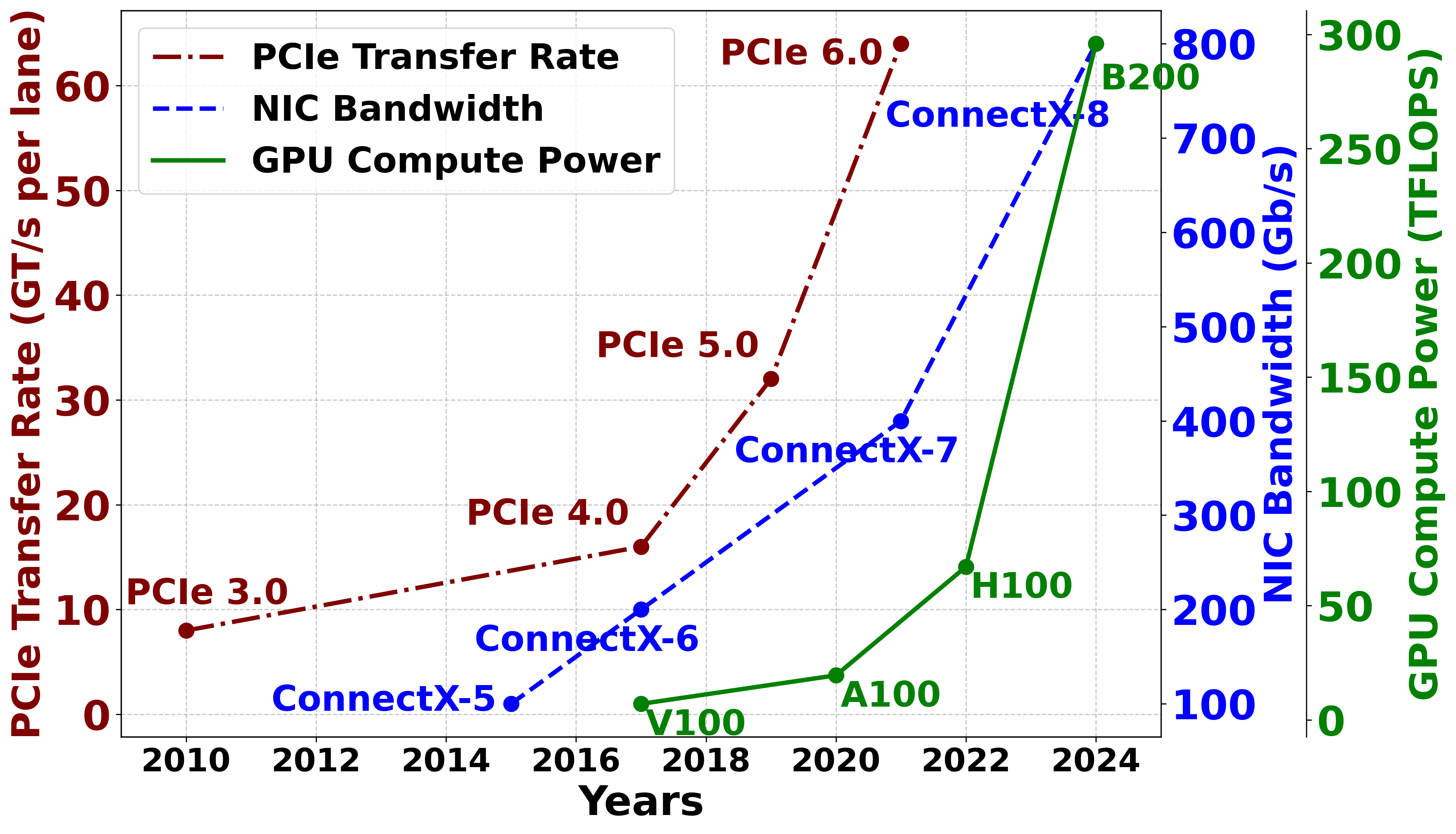}
    \caption{Evolution of AI cluster hardware.}
    \label{fig:evolution}
\end{figure}

Today, industries such as Meta~\cite{gangidi2024rdma}, and Alibaba~\cite{qian2024alibaba} train LLMs on specialized clusters equipped with homogeneous GPUs and high-speed Remote Direct Memory Access (RDMA) network interface cards (NICs). However, the assumption of the availability of homogeneous AI infrastructure may not always be true: (1) The hardware FLOPS and interconnect bandwidth are increasing at rates of 3.0$\times$ and 1.4$\times$ per year, respectively~\cite{gholami2024ai} (see \fref{fig:evolution}). Such rapid evolution poses financial challenges in maintaining a new generation of homogeneous GPU clusters; (2) Users of the shared cloud infrastructure may encounter significant queueing delays due to the unavailability of a fraction of the resource, to ensure homogeneity~\cite{motivationMSR-multitenant-cloud-training,motivation-alibaba-MLaaS,mo2024heet}. For example, Google Cloud comprises machine types with NVIDIA GPUs B200, H100, H200, A100, L4, T4, V100, P100, and P4~\cite{google-cloud-gpus}; and (3) Heterogeneity in interconnect bandwidth and latency on super-chip architectures such as the Grace-Hopper~\cite{motivation-onchip-hetero} results in performance variability. Consequently, there is a strong demand to leverage the heterogeneous compute and network infrastructure for LLM training during the transitional periods between device generations and within shared cloud infrastructure. 

Prior works~\cite{park2020hetpipe,jia2022whale,um2024metis,wu2025hetermoe, zhang2024hap,yang2024holmes} propose an optimal deployment plan for LLM training in a heterogeneous cluster comprising multiple GPU types and variable network bandwidth. The core idea of these solutions is to non-uniformly partition the workload across GPUs with different compute and interconnect (\eg, NVLink bandwidth and PCI interconnect generation) capabilities. To test the proposed prototypes for heterogeneous clusters, researchers rely on (a) real-world deployments~\cite{mei2025helix}, which are not scalable and accessible to all, or (b) analytical simulations~\cite{strati2025sailorautomatingdistributedtraining}, which do not mimic real-world conditions. 

Existing LLM training simulators, ASTRA-Sim and SimAI, lack features to simulate heterogeneity. For example, they do not support abstractions to specify custom device groups, workload distribution based on device capabilities, and matching model parameter shape prior to synchronization (details in~\S\ref{sec:design}).

\myparab{Our key idea.} We propose to design a heterogeneity-aware distributed training simulator framework that extrapolates the LLM infrastructure and accurately predicts the performance of the custom deployment. 

Our main contributions in this paper are as follows:
\begin{enumerate}
    \item We identify the requirements to build a heterogeneity-aware simulator (\S\ref{subsec:requirements}).
    \item We design abstractions and components on top of an existing simulator, SimAI~\cite{wang2025simai}, to satisfy the heterogeneity-aware simulator requirements. Our simulator design facilitates abstractions for custom (homogeneous and heterogeneous) device groups, hybrid parallelization strategies, and custom topologies. We implement system components such as non-uniform workload partitioning and resharding (~\S\ref{subsec:design}).
    \item We implement the abstractions required to simulate interconnect (or network) heterogeneity, \ie, NVswitch, NVLink, PCI, and NIC (network interface card) processing delays (\S\ref{subsec:design}).
    \item Based on our initial simulator prototype, we present the heterogeneity-aware compute and interconnect inferences (\S\ref{sec:eval}).
\end{enumerate}

\section{Distributed LLM training} \label{sec:motivation}

This section describes the specialized training infrastructure and parallelism techniques that improve the efficiency and scalability in distributed ML training.

\myparab{Parallelism techniques} 
Training large-scale models uses parallelism techniques such as Data parallelism (DP), Pipeline parallelism (PP), and Tensor parallelism (TP), to train a large volume of data in parallel, or to fit the model within the GPU's memory and compute limits. The model parameters, gradients, and activations are synchronized (commonly through AllReduce~\cite{thakur2003improving}) to guarantee uniform model updates across all GPU instances. Table~\ref{tab:parallelism-char} compares the communication patterns and the cost of each parallelism technique, which is the tradeoff that we pay for scalability if the communication cannot be hidden behind ongoing compute, \ie, {\em exposed communication}. In summary, DP synchronization involves large-sized flows, whereas TP requires high-frequency synchronization.

\begin{table}[t]
\centering
\scriptsize
\setlength{\tabcolsep}{4.7pt}
\renewcommand{\arraystretch}{1.2}
\begin{tabularx}{\linewidth}{|c|c|c|c|c|}
\hline
\multicolumn{2}{|c|}{\textbf{Attribute}} & \textbf{DP} & \textbf{TP} & \textbf{PP} \\
\hline
\multirow{2}{*}{Exposed Comm} & Forward & No & Yes & Yes \\ \cline{2-5}
& Backward & Yes & No & Yes \\
\hline
\multicolumn{2}{|c|}{Frequency (per iteration)} & $2$ (low) & $350$ (high)  & $8$ (moderate) \\
\hline
\multicolumn{2}{|c|}{Avg. comm size (per collective)} & $4.4GB$ (large) & $67KB$ (small) & $67KB$ (small)\\
\hline
\end{tabularx}
\caption{Exposed communication characteristics of LLM parallelism techniques in distributed training illustrated using the Llama-2 70B~\cite{aws2025llama2tutorial, touvron2023llama} model where DP, TP and PP is 32, 8 and 8 respectively and total number of GPUs are 2048.}
\label{tab:parallelism-char}
\end{table}

\eat{Table~\ref{tab:parallelism-char} summarizes the exposed communication characteristics of common parallelism strategies—Data Parallelism (DP), Tensor Parallelism (TP), and Pipeline Parallelism (PP)—used in large language model (LLM) training. It highlights key differences in communication visibility, frequency, and volume. DP primarily exposes communication during the backward pass through gradient synchronization, with relatively low frequency but large communication sizes due to full model replication. TP, in contrast, requires fine-grained communication during the forward pass to exchange partitioned tensor slices, resulting in high frequency but small-sized messages. PP introduces communication at both forward and backward stages to transfer activations and gradients between pipeline stages, with moderate frequency and small collective sizes. These distinctions are critical for optimizing interconnect usage and selecting appropriate parallelism strategies based on system architecture and model scale.}

\begin{figure}[t]
    \centering
    \includegraphics[width=1\linewidth]{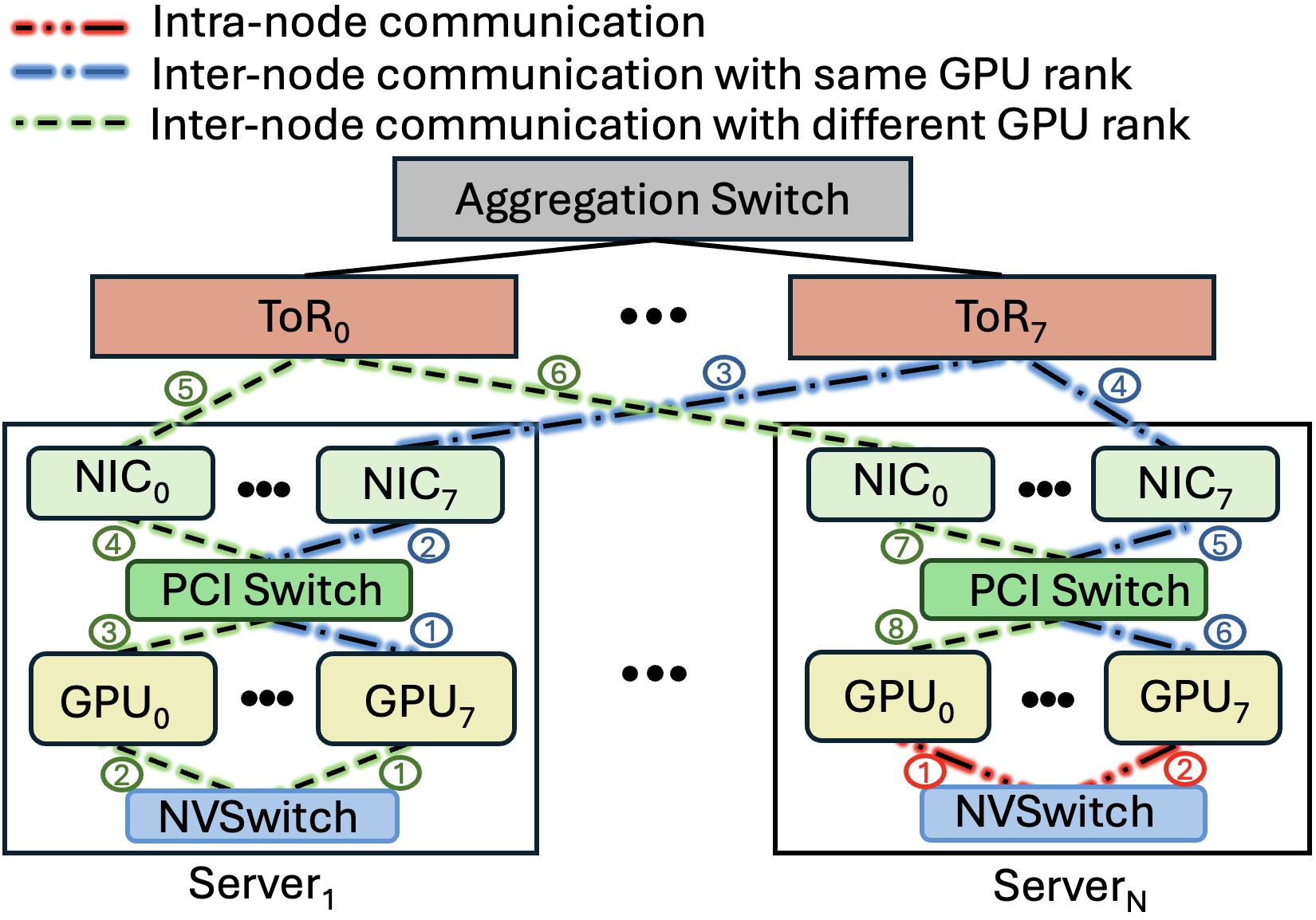}
    \caption{Rail-only topology for LLM training. (a) Intra-node communication ($Server_N:GPU_0$ to $Server_N:GPU_7$), (b) Inter-node communication with same local GPU rank ($Server_1:GPU_7$ to $Server_N:GPU_7$) (c) Inter-node communication with different local GPU rank ($Server_1:GPU_7$ to $Server_N:GPU_0$)   }
    \Description{rail-only-topology}
    \label{fig:training-infra}
\end{figure}

\myparab{Large model training infrastructure}
A typical data center training infrastructure comprises 10s of thousands of GPUs distributed across physical machines (viz, node), each with {\em eight} homogeneous GPUs. Each GPU is assigned a {\em global rank}, which is unique across the cluster, and a {\em local rank}, which is unique within the node. The local and global ranks help to indicate the GPU's role in intra-node and inter-node communication for model synchronization.  For example, in DP, each local rank handles a different data subset, whereas in TP, each local rank handles a different model partition. The global rank is used to determine the communication path in global AllReduce operations. Figure \ref{fig:training-infra} illustrates the key aspects of the Rail-only topology~\cite{wang2024rail} using {\em three} example communications. Frequent communication must be mapped within a node to leverage high bandwidth, low-latency NVLink (\eg, TP). Inter-node communication uses dedicated NICs via dedicated PCI channels to improve the collective communication efficiency. Rail-only design eliminates the need to send traffic through the aggregation switches, thus accelerating training.

\begin{table}[t]
 
  \setlength{\tabcolsep}{2pt}
  \scriptsize
  \begin{tabular}{|l|c|c|c|c|c|}
    \hline
    \textbf{Features} & \makecell{\textbf{Chakra~\cite{sridharan2023chakra}+} \\  \textbf{AstraSim~\cite{won2023astra}}} &  \makecell{\textbf{SimAI}  \\ \textbf{~\cite{wang2025simai}}} & \makecell{\textbf{Echo} \\ \textbf{~\cite{feng2024echo}}} &  \makecell{\textbf{HTsim} \\ \textbf{~\cite{htsim}}} & \makecell{\textbf{Sailor} \\ \textbf{~\cite{strati2025sailorautomatingdistributedtraining}}} \\
    \hline
    \makecell[l]{Support for any ML model} & \cmark & \xmark & \xmark & -- & \cmark \\
    \hline
    Trace extrapolation support & \xmark & \cmark & \cmark & -- & \cmark \\
    \hline
    \makecell[l]{Full stack training simulation} &  \cmark & \cmark & \xmark & \xmark & \xmark \\
    \hline
    Collective optimization support & \xmark & \cmark & \cmark & -- & \xmark \\
    \hline
    \makecell[l]{Network protocol simulation} & \cmark & \cmark & \xmark & \cmark & \xmark \\
    \hline
    \makecell[l]{Heterogeneous cluster simulation} & \xmark & \xmark & \xmark & \xmark & \cmark \\
    \hline
  \end{tabular}
  \caption{Comparison of SOTA LLM training simulators}
  \label{tab:simulator-comparison}
\end{table}

\myparab{Existing LLM training simulators.} 
The state-of-the-art (SOTA) simulators (see Table~\ref{tab:simulator-comparison}) can be classified as: (1) full-stack training simulators~\cite{won2023astra,wang2025simai,feng2024echo}, and (2) network simulators~\cite{htsim,bai2024unison}. 
Chakra\cite{sridharan2023chakra} and AICB~\cite{aicb_benchmark} executes the real-world training workload on the GPUs to capture the dependencies between compute, communication, and memory operations to generate the workload file. 
SimAI~\cite{wang2025simai},  Echo~\cite{feng2024echo}, and Sailor~\cite{strati2025sailorautomatingdistributedtraining} extrapolate the real-world traces obtained for a subset of the GPU cluster. However, Echo and Sailor use analytical modeling for network simulation.
ASTRA-sim and SimAI simulate the entire training stack, for example, workload generation and partitioning, communication optimizations, and simulating training events over the training cluster. HTSim~\cite{htsim} and ns3~\cite{bai2024unison} are network simulators that support data center network topologies and transport protocols specialized for ML workloads (\ie, RoCE). ns3 models "protocol-true" network communication events. HTSim ignores the protocol implementation and only models the communication time, leading to significant simulation speedup. The SOTA simulators do not support the key ingredients for LLM training in heterogeneous infrastructure, such as non-uniform workload partitioning, resharding model parameters, and the abstractions for custom cluster, framework, and topology (see~\S\ref{subsec:requirements}).

\section{Heterogeneity-aware LLM training} \label{sec:heter-aware}

Real-world training clusters often feature variability in GPU compute capacity, communication bandwidth, and network delays, causing synchronization bottlenecks that result in suboptimal training performance~\cite{jia2022whale,yan2025hexiscaleaccommodatinglargelanguage}. To address the performance challenge, SOTA heterogeneity-aware LLM training solutions~\cite{um2024metis,park2020hetpipe,yang2024holmes} implement ideas such as (a) non-uniform workload partitioning, based on compute capabilities, \eg, the LLM model's MLP layer is compute-intensive and can be assigned to high compute GPUs for speedup~\cite{wu2025hetermoe}, (b) heterogeneity-aware placement of distributed LLM model slices, \ie, layers and tensors~\cite{yan2025hexiscaleaccommodatinglargelanguage}. For example, leverage high bandwidth interconnects for model slices that communicate large amounts of data frequently, and (c) heterogeneity-aware training data sharding and orchestration, \eg, in the case of multimodal data, image/video data must be trained on high-speed hardware~\cite{wang2025spindle}.

\begin{figure}[t]
    \centering
    \includegraphics[width=0.8\linewidth]{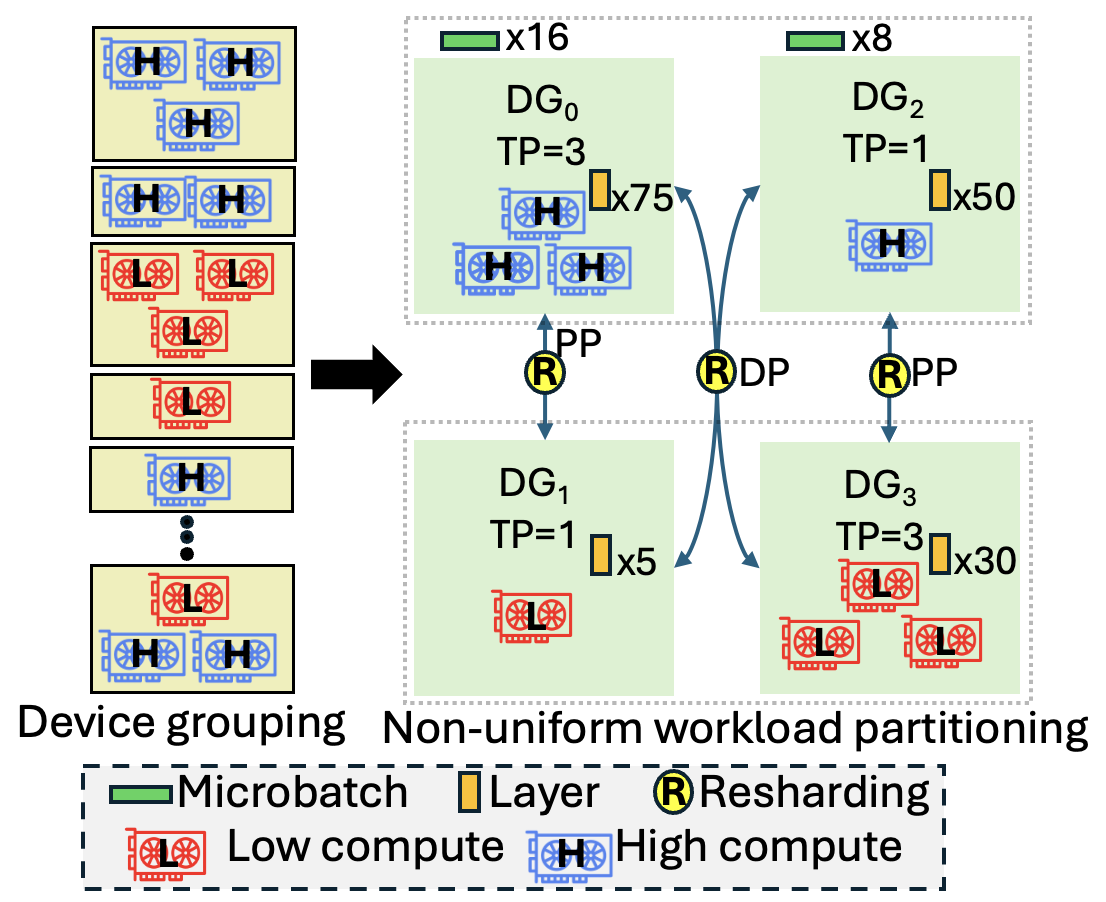}
    \caption{Example workflow of a heterogeneity-aware distributed training solution. 
    }
    \label{fig:het-aware-abstractions}
\end{figure}

\fref{fig:het-aware-abstractions} shows a heterogeneous cluster configuration for training the Llama-2 (70B)~\cite{touvron2023llama} model comprising $80$ layers. The cluster comprises two nodes, $Node_A$ and $Node_B$. $Node_A$ is equipped with $4 \times H100$ 80G (in blue) connected via NVLink (600GB/s) and ConnectX-6 (200 Gb/s), whereas $Node_B$ is equipped with $4 \times A100$ 40G GPUs (in red) connected via NVLink (900 GB/s) and Intel E830-CQDA2 (200 Gb/s) (see  table~\ref{tab:gpu_specific_delay_table} for detailed specs). Assume that the global batch size is $24$, and the microbatch size is $1$. The LLM training workflow, given the above-described heterogeneous infrastructure, comprises (a) device group generation, consisting of homogeneous and heterogeneous GPUs, (b) workload partitioning based on the device group's capabilities, and (c) resharding the model parameters to enable synchronization across non-uniform partitions.

\myparab{Device grouping.} A {\em device group (DG)} refers to a collection of GPU nodes that divide the model for a given batch size to form a pipeline (a.k.a., {\em PP group}). A {\em TP group} comprises GPUs that jointly compute one copy of the model partition. A {\em DP group} includes all the GPUs used to train the model. \fref{fig:het-aware-abstractions} shows the device group combinations with $4 \times H100$ (denoted as H) and $4 \times A100$ (denoted as A) GPUs. For example, (H,H,H), (H,H), (A,A,A), (A), (A,A), (A,H,H) and so on. 

\myparab{Non-uniform workload partitioning.} SOTA heterogeneity-aware solutions generate all possible combinations of: (a) device groups, (b) hybrid parallelism strategy (\ie, combining PP, TP, or DP) with varying parallelism degree, and (c) non-uniform partitioning of data (DP), layers (PP), and tensors (TP). 
The SOTA solutions aim to improve resource utilization and speed up training time by mapping the device groups (and devices) with the hybrid parallelism technique and degree, appropriately.   

\fref{fig:het-aware-abstractions} illustrates the use of hybrid parallelism strategies. We observe a non-uniform batch size distribution (DP) and tensor distribution (\ie, TP=1 and TP=3). To exploit heterogeneity, (1) We assign more transformer layers to high compute GPUs (\ie, 75 in  $DG_{0}$ and 50 in $DG_{2}$),(2) Variable TP degrees assigned based on the GPU count per device group (\ie, TP=1 and TP=3), (3) To balance the computation speeds in a heterogeneous setting, non-uniform batch sizes are assigned to each device group in DP, for example, $16$ on the high compute DP device groups and $8$ on the low compute device groups.

\begin{table}[t]
\centering
\scriptsize
\setlength{\tabcolsep}{1pt}
\renewcommand{\arraystretch}{1.2}
\begin{tabular}{|c|>{\centering\arraybackslash}p{1cm}|>{\centering\arraybackslash}p{1cm}|>{\centering\arraybackslash}p{1cm}|c|}
\hline
\multirow{2}{*}{\textbf{Heterogeneity-aware SOTA}} & \multicolumn{3}{c|}{\textbf{Parallelism strategy}} & \multirow{2}{*}{\shortstack{\textbf{Need for} \\ \textbf{Resharding}}}  \\
\cline{2-4}
 & \textbf{DP} & \textbf{TP} & \textbf{PP} & \\
\hline
Hexiscale~\cite{yan2025hexiscaleaccommodatinglargelanguage}, Metis~\cite{um2024metis}, Whale~\cite{jia2022whale}  & \cmark & \cmark & \cmark & \cmark \\

\hline
PipePer~\cite{zhang2023pipepar}, HeterMoE~\cite{wu2025hetermoe} HetPipe~\cite{park2020hetpipe}  & \xmark & \xmark & \cmark & \xmark \\
\hline
HAP~\cite{zhang2024hap}     & \xmark & \cmark & \xmark & \cmark  \\
\hline
HetSeq~\cite{ding2021hetseq}     & \cmark & \xmark & \xmark & \cmark  \\
\hline
\end{tabular}
\caption{Comparing optimization strategies and overheads incurred by heterogeneity-aware SOTA.}
\label{tab:hetero-paper-comparison}
\end{table}

\myparab{Resharding.}
 The synchronization process during the backward pass must be preceded by resharding to match the parameter shape of the synchronizing device groups. Resharding is inevitable under the following conditions: (1) the microbatch size processed by the source DP group is different from that of the destination DP group, and (2) the TP degree between the communicating nodes is not uniform. 
In our example (see~\fref{fig:het-aware-abstractions}), for each inter-node communication, at least one of these conditions holds, making it essential to reshard the parameters prior to a collective operation. The variation in model layer distribution (PP) by itself does not require resharding since the communication is sequential.
Table~\ref{tab:hetero-paper-comparison} compares the SOTA heterogeneity-aware solutions. Despite the use of non-uniform DP, HetPipe~\cite{park2020hetpipe} does not require resharding because it uses a parameter server for synchronization. 

\section{Design and implementation} \label{sec:design}

We discuss the design and the initial prototype implementation of a heterogeneity-aware simulation framework for LLM training. To achieve this, we extend SimAI~\cite{wang2025simai}, which originally supports homogeneous GPU clusters and interconnects, to incorporate the abstractions and components.

\subsection{ Design requirements} 
\label{subsec:requirements}

\begin{table}[t]
\centering
\small
\renewcommand{\arraystretch}{1.2}
\resizebox{\linewidth}{!}{%
\begin{tabular}{|l|l|c|c|c|c|}
\hline
\multicolumn{2}{|c|}{\textbf{Key idea of heterogeneity-aware SOTA}} & \textbf{C1} & \textbf{C2} & \textbf{C3} & \textbf{C4} \\
\hline
\multirow{2}{*}{\parbox[c][4.5em][c]{1.8cm}{\centering Non-uniform workload \\partitioning}} 
& Uses TP, DP, and PP~\cite{jia2022whale,um2024metis,yan2025hexiscaleaccommodatinglargelanguage} & \cmark & \cmark & \cmark & \cmark\\ \cline{2-6}
& Uses PP~\cite{zhang2023pipepar,wu2025hetermoe,park2020hetpipe} & \cmark & \xmark & \cmark & \cmark \\ \cline{2-6}
& Uses TP~\cite{zhang2024hap}  & \cmark & \cmark & \cmark & \cmark \\
\hline
\multicolumn{2}{|l|}{Exploits network heterogeneity~\cite{li2024tccl,yang2024holmes}} & \xmark & \xmark & \cmark & \cmark \\
\hline
\multicolumn{2}{|l|}{Exploits heterogeneity in training data (Multimodal data)~\cite{wang2025spindle}} & \cmark & \cmark & \xmark & \xmark \\
\hline
\end{tabular}
}
\caption{Components required for simulating heterogeneity-aware SOTA solutions. }

\label{tab:sim-req}
\end{table}

We identify the gaps with SOTA simulators (\S\ref{sec:motivation}), and present the abstractions ({\em A}) and components ({\em C}) needed for a heterogeneity-aware simulator design. Table~\ref{tab:sim-req} summarises the support required from a simulator to test the heterogeneity-aware SOTA solutions.

\myparab{[A1] Abstractions for custom device groups and hybrid parallelism strategies.}
Existing simulators~\cite{wang2025simai,won2023astra} do not have the provision to specify custom device groups, \ie, the set of heterogeneous compute devices that process a specific model slice (model partition or a set of model layers) or a training data slice. 
To enable non-uniform workload partitioning and heterogeneity-aware placement, the LLM training simulator must additionally support abstractions for: (a) heterogeneity-aware device groups across nodes, (b) custom parallelism configurations (Pipeline parallelism (PP), Tensor parallelism (TP), and Data parallelism (DP)) with variable batch sizes, and (c) flexible mapping of parallelism strategies to device groups.

\myparab{[A2] Custom cluster and topology specification.}
The heterogeneity-aware simulator must provide abstractions to define 
diverse interconnects (\eg, PCIe, NVswitch, and NVLink), bandwidth/latency parameters, and topological arrangements, enabling accurate simulation of diverse hardware infrastructures.

\myparab{[C1] Non-uniform workload partitioning.}
The heterogeneity-aware simulator must: (a) distinguish between GPU types (\eg, A100 vs. H100) and generate distinct workload traces tailored to the device group’s role in the parallelism strategy, and (b) correctly simulate the non-uniform hybrid parallelism (\ie, PP, TP, and DP) and the collective communication over custom device groups.

\myparab{[C2] Resharding support for shape mismatch.}
Training across heterogeneous device groups with non-uniform parallelism configurations may result in tensor (or activation) shape mismatch during synchronization. 
Suppose Llama-2 (70B)~\cite{touvron2023llama} with 80 layers is trained on heterogeneous GPUs with two device groups (DG). $DG_1$ performs TP on 75 layers over 3 GPUs, and $DG_2$ performs TP on 5 layers over 2 GPUs. Due to the mismatch between the tensor shapes of $DG_1$ and $DG_2$, resharding is needed.
The simulator must support automated resharding to adjust tensor shapes and ensure correctness in collective communication operations.

\myparab{[C3] Heterogeneity-aware collective communication.}
Existing LLM training simulators~\cite{wang2025simai} imitate  NCCL~\cite{nccldoc} collective communication library optimizations. NCCL optimizes collective communication for: (a) efficient intra-node communication using bandwidth-aware graphs, (b) detects and selects efficient inter-node transport, 
and (c) maps logical ranks to physical devices to optimize performance.
However, NCCL assumes GPU homogeneity and works only for NVIDIA GPUs. A heterogeneity-aware collective communication must support: 
(a) Graph generation for efficient collective communication in a heterogeneous cluster, \ie, clusters with CPU-only nodes, or asymmetric architecture (\eg, CPU+GPU+NPU), and (b) must be vendor agnostic.

\myparab{[C4] Heterogeneous compute and interconnect simulation.}
The simulator must accurately measure and simulate the compute performance based on the bottleneck device in the ongoing transaction, and simulate custom network topology, interconnect capacities, and their delays.

\subsection{Design overview} \label{subsec:design}
\begin{figure}
    \centering
    \includegraphics[width=0.9\linewidth]{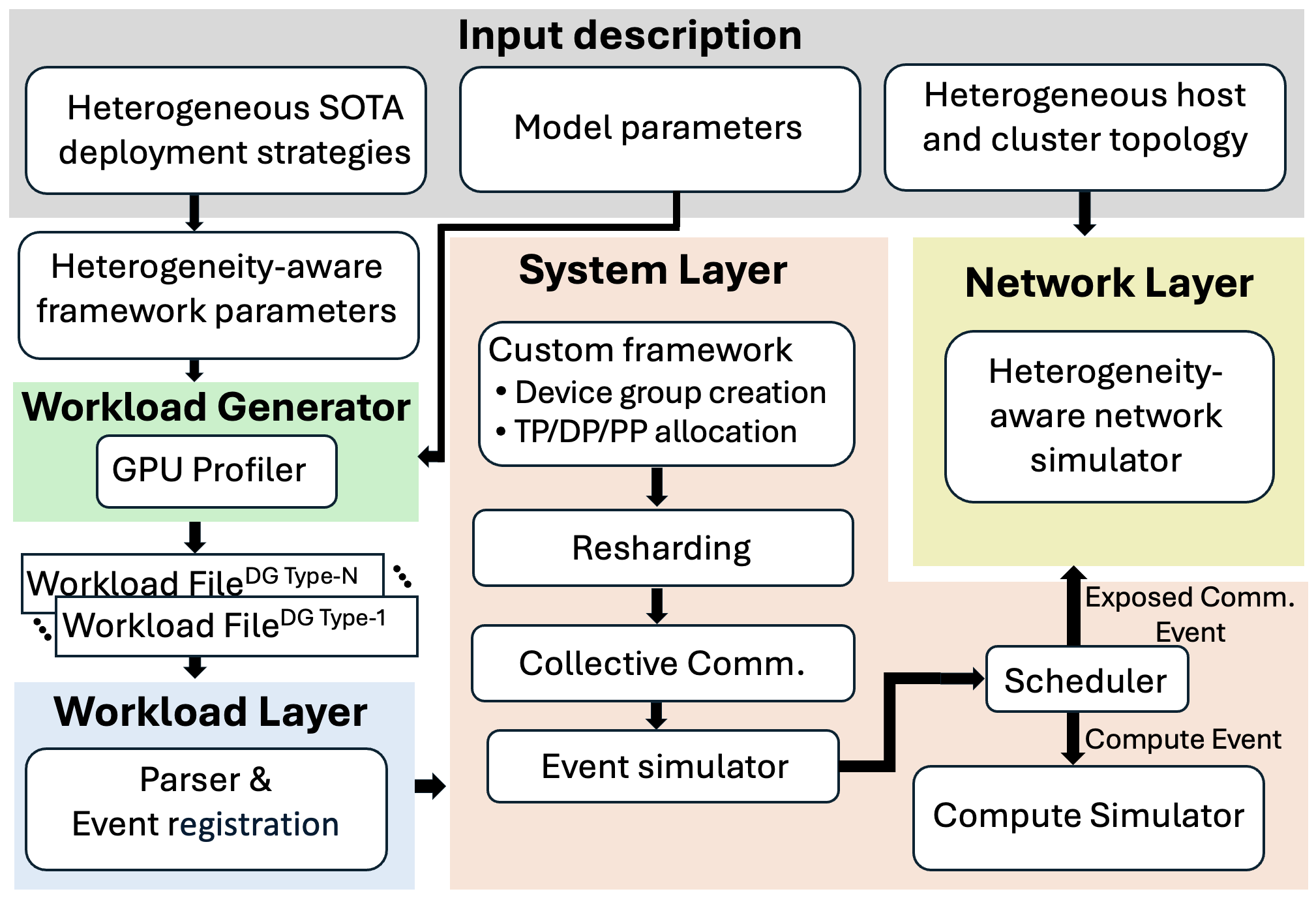}
    \caption{Design of heterogeneity-aware LLM training simulator.}
    \label{fig:design}
\end{figure}

\fref{fig:design} shows the primary components of our simulation framework.

\myparab{Input Description {\bf[A1, A2]}.} \label{subsec:input-dec} 
The input description component allows the user to feed in the custom heterogeneous configurations. The abstraction allows the user to specify: (1) {\em Model parameters.} (see table~\ref{tab:model_spec}) such as model dimensions, layers, and training hyperparameters; 
(2) {\em Framework parameters.} comprises (a) custom device group specification, (b) parallelism degree for PP, TP, and DP, given non-uniform workload partitioning, and (c) custom parallelism to device group mapping. A device group is represented as, $DG_1=\{(GPU_{type1}, count_1), ..., (GPU_{typeN}, count_N)\}$; (3) {\em Heterogeneous host and cluster topology.} (see table~\ref{tab:gpu_specific_delay_table}) User describes the heterogeneous topology, compute and interconnect capacities, detailing their latency and bandwidth metrics, which allows the simulator to model diverse network behavior.

\myparab{Workload layer {\bf[C1]\eat{(partial)}}.}
Based on the input description, the workload generator profiles the workload layers on the specified device (\ie, A100, H100) to generate workload files specific to device groups.
We implemented a custom parser that registers the compute and communication events based on the device group's workload file.

\myparab{System layer {\bf[C1, C2, C3] }.}
The system layer is primarily responsible for logical resource management and scheduling.
Using the framework parameters, the system layer creates custom device groups comprising homogeneous or heterogeneous GPUs. The device groups are then mapped to the tensor and data parallelism based on the mentioned degree of parallelism. In the case of non-uniform TP/DP groups (e.g., TP=3 ($DG_0$) communicating with TP=1 ($DG_1$) in ~\fref{fig:het-aware-abstractions}), we need to reshape the parameters and register the communication event for parameter synchronization over the custom channel. The event simulator queues the registered events and maintains event logs over the distributed execution timeline. The scheduler coordinates the event stream between the compute and network simulators, and ensures accurate modeling of event dependencies, resharding delays, and bandwidth contention. 
For optimized collective communication, the CCL must generate logical topology graphs based on the heterogeneous cluster capabilities. 

\myparab{Network layer {\bf[C4]}.}
We design a heterogeneity-aware network simulator engine on top of SimAI's ns3 module~\cite{bai2024unison}. 
The simulator uses the input description to instantiate and configure custom devices, interconnects, and topology (see table~\ref{tab:gpu_specific_delay_table}).We simulate delays in PCIe and NVLink specific to the custom interconnects to represent intra-node and inter-node communication traits. In the future, we plan to emulate fluctuating NIC bandwidth and processing delays to mimic factors such as queue management.
This helps precise modeling of collective operations and uncovers the impact of heterogeneity on the overall completion time of LLM model training. 
In our prototype, we modified the SimAI's NS-3 component responsible for simulating RDMA packet transmission (\ie, the {\em QbbChannel} module) to incorporate the delays specific to different interconnects.

\section{Evaluation} \label{sec:eval}
\begin{table*}[t]
\centering
\scriptsize
\resizebox{\textwidth}{!}{%
\begin{tabular}{|c|c|c|c|c|c|c|c|}
\hline
\textbf{Architecture} & \textbf{GPU}  & \textbf{\makecell{NVLink BW\\(Gbps)}} & \textbf{\makecell{NVLink delay\\(ns)}} & \textbf{\makecell{PCIe BW\\(Gbps)}} & \textbf{\makecell{PCIe latency\\(ns)}} & \textbf{\makecell{NIC BW\\(Gbps)}} & \textbf{\makecell{NIC processing delay\\(ns)}} \\
\hline
\makecell{Ampere~\cite{connectx6,nvidia_a100}} &\makecell{ A100 (40GB)} & \makecell{4800 (Gen 3)} & 30.66 & \makecell{512 (Gen 4)} & \makecell{$2\times287.5$} & \makecell{200 (ConnectX-6)} & 368 \\
\hline
\makecell{Hopper~\cite{intelE830CQDA2,nvidia_h100} }&\makecell{ H100 (80GB)} & \makecell{7200 (Gen 4)} & 20.44 & \makecell{1024 (Gen 5)} & \makecell{$2\times143.75$} & \makecell{200 (Intel E830-CQDA2) } & 368 \\
\hline
\end{tabular}
}

\caption{Cluster specific configurations}
\label{tab:gpu_specific_delay_table}
\end{table*}

\begin{table*}[t]
\centering
\scriptsize
\resizebox{\textwidth}{!}{%
\begin{tabular}{|c|c|c|c|c|c|c|c|c|c|c|c|c|c|c|c|c|}
\hline
\multicolumn{8}{|c|}{
\textbf{Model Characteristics}} & \multicolumn{6}{c|}{\textbf{Deployment Characteristics}} \\
\hline
\textbf{Model} & \textbf{\makecell{Num. of\\Layers}} & \textbf{\makecell{Hidden\\Size}} & \textbf{\makecell{Num. of\\Attention\\Heads}} & \textbf{\makecell{FFN\\Hidden\\Size}} & \textbf{\makecell{Sequence\\Length}} & \textbf{\makecell{Max\\Positional\\Embeddings}} & \textbf{\makecell{Vocab\\Size}} & \textbf{\makecell{World\\Size}} & \textbf{TP} & \textbf{PP} & \textbf{DP} & \textbf{\makecell{Global\\Batch\\Size}} & \textbf{\makecell{Micro\\Batch\\Size}} \\
\hline
GPT 6.7B~\cite{LanguageModelsareFew-ShotLearners,TransformerArchitectureMetricsCheatSheetGPT3,llm_training_on_gpu_clusters,radford2019language} & 32 & 4096 & 32 & 16384 & 2048 & 2048 & 50257 & 128 & 4\ & 1 & 32 & 976 & 8 \\
\hline
GPT 13B~\cite{LanguageModelsareFew-ShotLearners,TransformerArchitectureMetricsCheatSheetGPT3,llm_training_on_gpu_clusters,radford2019language} & 40 & 5120 & 40 & 20480 & 2048 & 2048 & 50257 & 256 & 8 & 1 & 32 & 976 & 8 \\
\hline
Mixtral 8x7B~\cite{aicb_benchmark,chinesemixtral,mixtralhuggingface,jiang2024mixtral,llm_training_on_gpu_clusters,radford2019language} & 32 & 4096 & 32 & 14336 & 2048 & 131072 & 32000 & 128 & 2 & 1 & 64 & 1152 & 4 \\
\hline
\end{tabular}
}
\caption{Model specifications}
\label{tab:model_spec}
\end{table*}

\begin{figure*}[t]
    \centering
    \includegraphics[width=\textwidth]{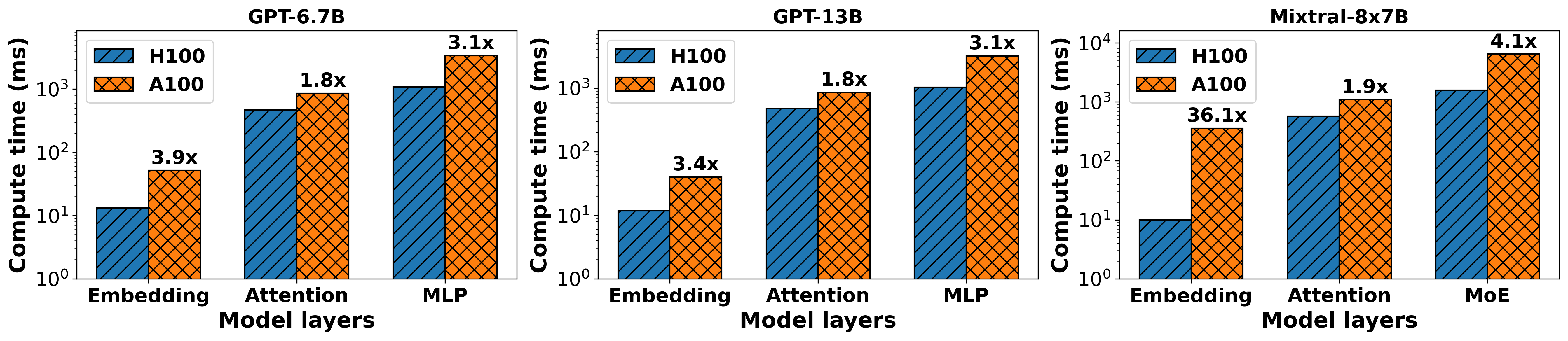}
    \caption{Per-layer compute time for GPT6.7B, GPT13B and Mixtral-8x7B for one iteration across H100 and A100 GPUs.}
    \label{plot:per-layer-compute}
\end{figure*}

\begin{figure*}[t]
    \centering
    \includegraphics[width=\textwidth]{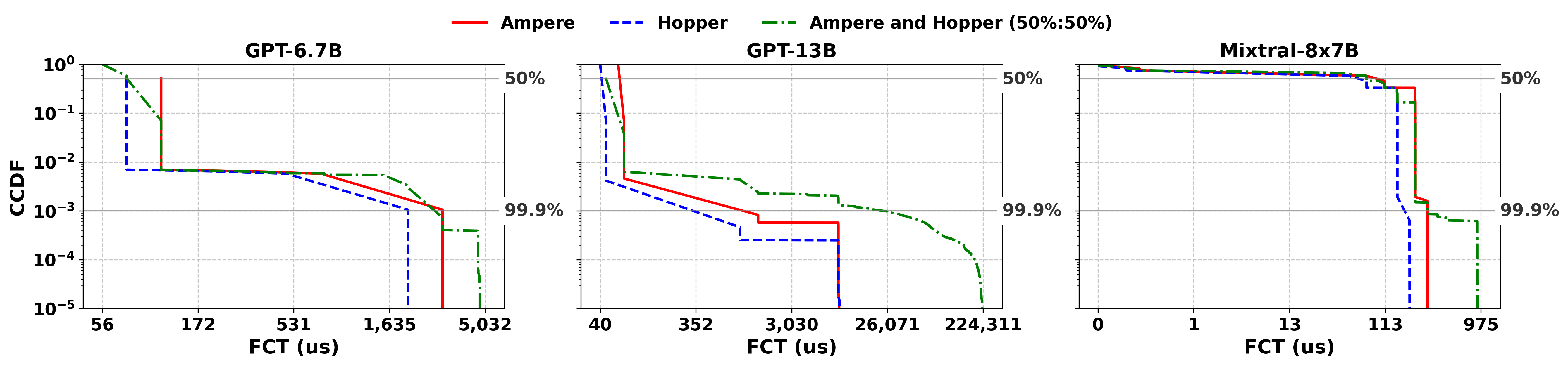}
    \caption{FCT distribution for GPT-6.7B, GPT-13B and Mixtral-8x7B for one iteration across homogeneous and heterogeneous GPU clusters with Ampere and Hopper configurations. 50\% and 99.9\% represent median and 99.9 percentile FCT, respectively.}
    \label{plot:fct}
\end{figure*}

Using our initial simulator prototype, we demonstrate the impact of heterogeneity on training real-world LLM models.

\myparab{Setup.}
We set up the SimAI simulation framework \cite{wang2025simai} on a 32-core AMD EPYC-9354 node with A100 and H100 GPUs. 
All the experiments were run for {\em one} training iteration 
, unless mentioned otherwise.

\myparab{Workload details.}
We generate the model workloads using the AICB workload benchmark \cite{aicb_benchmark}, on A100 and H100 GPUs, to run simulations for homogeneous and heterogeneous configurations.
We simulate heterogeneity over {\em three} LLM models, {\em GPT-6.7B}, {\em GPT-13B}, and {\em Mixtral 8x7B} with realistic training configurations (see table \ref{tab:model_spec}). This choice was made to include models with different architectures, decoder-based (\eg, GPT) and MoE models (\eg, Mixtral), and model sizes.

\myparab{Interconnect configuration.}
We simulate a rail-only topology, with each node having 8 GPUs and 8 RoCE NICs.
Table \ref{tab:gpu_specific_delay_table} shows the compute and interconnect characteristics in terms of bandwidth and latency. For example, to simulate an Ampere configuration, with A100 (40 GB) cluster, ConnectX-6 NICs (200 Gbps), PCIe Gen4 (512 Gbps), and NVLink Gen3 (4800 Gbps), we compute the PCIe and NVLink delays using the formula, 
$\text{delay} = \frac{\text{Jumbo frame size(in bytes) x 8}}{\text{uni-directional bandwidth (Gbps)}}$ , considering a jumbo frame size of 9200 bytes \cite{jumbo_mtu}. 
For inter-node GPU communication, the PCIe latency involves two trips: (1) GPU to PCIe Switch, and (2) PCIe Switch to NIC.  

\myparab{Metrics.}
For heterogeneous compute experiments, we measure the compute time for the model layers. 
{\em Flow completion time (FCT)} is used to measure the time taken by each flow (5-tuple) during the collective communication, such as AllReduce. 

\myparab{[Q1] Layer-wise compute time across GPU generations.} 
\fref{plot:per-layer-compute} shows the compute time for the model layers, {\em Embedding, Attention, and MLP (or MoE)}, for one iteration across different GPU generations and LLM models. 
We observe that the compute time degradation for {\em MLP} layer when processed by low-compute A100 GPUs ranges between $3\times$---$4\times$, whereas the degradation of the {\em attention} layer is up to $1.9\times$ across all models. 

To leverage heterogeneous compute, one may decide to assign an MLP layer to high compute GPUs for speedup~\cite{wu2025hetermoe}. Despite the significant compute time degradation for {\em embedding} layer ($36.1X$), it is not a good candidate for optimization since this layer is processed only once within an iteration. Such a heterogeneity-aware simulation helps the LLM training deployer make an informed choice while exploiting heterogeneity.

\myparab{[Q2] Impact of interconnect heterogeneity on FCT of all collective operations.}
Figure \ref{plot:fct} shows the FCT distribution (CCDF) for collective operations within and across the server nodes in one iteration. Given that the system layer is partially implemented, the {\em Ampere and Hopper} configuration refers to only the interconnect simulation. Since collective communication is a blocking operation, the flow with the highest FCT value determines the bottleneck flow for one training iteration. 
We observe that the performance degradation for the heterogeneous infrastructure, \ie, Ampere and Hopper (50\%:50\%) in comparison to the Ampere configuration is up to 9\% for GPT-6.7B model, 2428 \% (25.3X) for GPT-13B, and 0.4\% for Mixtral 8x7B.
Incorporation of compute heterogeneity along with optimizations provided by heterogeneous training solutions (like Metis\cite{um2024metis}) would lead to decreased tail FCTs for heterogeneous clusters.

\section{Related work}

\myparab{Full-stack simulators for distributed training.}
ASTRA-sim~\cite{won2023astra} and Meta's Arcadia~\cite{Arcadia} model the parallelization strategies, scheduling policies, compute and memory design, and the network layer (transport layer and network end-point). However, they rely on realistic workload traces and fail to generate real-world-like communication patterns. 
On the contrary, SimAI~\cite{wang2025simai} extrapolates small-scale, realistic workload for large training clusters, and mimics the collective communication optimizations, but fails to account for overlapping communication with computation time, and Echo~\cite{feng2024echo} fills this gap. However, all prior works assume a homogeneous distributed training infrastructure.

\myparab{Optimizations in the presence of heterogeneous distributed training infrastructure.}
To optimize training completion time in a heterogeneous infrastructure setting, recent works~\cite{park2020hetpipe,ye2024asteroid,um2024metis,jia2022whale,zhang2023pipepar,wu2025hetermoe,yan2025hexiscaleaccommodatinglargelanguage,yang2024holmes} propose to leverage non-uniform degrees of TP, DP \& PP. Few prior works address compute and network heterogeneity~\cite{park2020hetpipe,ye2024asteroid,um2024metis,jia2022whale,zhang2023pipepar,yan2025hexiscaleaccommodatinglargelanguage,zhang2024hap,wu2025hetermoe}, whereas other works~\cite{yang2024holmes,li2024tccl} address only network heterogeneity.Some of these works test their solution on a real-world testbed, while others use an analytical approach. Our work complements the aforementioned solutions by proposing heterogeneity-aware abstractions that assist in simulating state-of-the-art solutions.

\myparab{Heterogeneity-aware simulators.}
Few works~\cite{mei2025helix,cho2024llmservingsim} simulate the distributed inference workload, whereas Sailor~\cite{strati2025sailorautomatingdistributedtraining} uses the analytical approach for distributed training for heterogeneous infrastructure. However, all these works miss out on realistic optimizations such as communication and computation overlap.
We fill this gap with a heterogeneity-aware, full-stack simulator for LLM training.

\section{Conclusion} \label{sec:discussion}
We identify the requirements to design a heterogeneity-aware LLM training simulator, and design the necessary abstractions and components to fill the gap in state-of-the-art LLM simulators. Our evaluation shows that an LLM training deployer can draw inferences from our simulator and plan an optimal deployment for a heterogeneous infrastructure. As part of the future work, we plan to extend this work to support a heterogeneity-aware LLM inference simulator.


\bibliographystyle{ACM-Reference-Format}
\bibliography{reference}


\end{document}